\documentstyle[12pt,epsfig]{article}

\begin{document}

\title{Born-Infeld generalization of the Reissner-Nordstr\"om black hole}

\author{Nora Bret\'on  \\
{\it Departamento de F\'{\i}sica, CINVESTAV del IPN.}\\
 {\it Apartado 14--740, 07000. M\'exico, D. F.} \\}

\maketitle
\begin{abstract}
In this work we study the trajectories of test
particles in a geometry that is the nonlinear electromagnetic
generalization of the Reissner-Nordstr\"om solution. The studied
spacetime is a Einstein-Born-Infeld solution, nonsingular outside a
regular event horizon and  characterized by three
parameters: mass $M$, charge $Q$ and the Born-Infeld parameter $b$
related to the magnitude of the electric field at the origin.
Asymptotically it is a Reissner-Nordstr\"om solution.
\end{abstract} 

\section{Introduction}

The Reissner-Nordstr\"om (R-N) solution (characterized by its charge $Q$
and mass $M$) turns out to be the final fate of a charged star, having as
uncharged limit the Schwarzschild black hole, then it is of interest to
investigate in more detail its nonlinear electromagnetic generalization. 

Moreover, Einstein-Maxwell (EM) theory is a consistent truncation of the
supergravity. Any asymptotically flat solution of EM theory with an
extremal charge and vanishing angular momentum defines a BPS solution of
supergravity \cite{BPS}.In particular, extremal ($Q=M$) 
Reissner-Nordstr\"om black hole is a BPS state in supersymmetry. States
that are slightly out of BPS limit have interest in the study of black
hole horizons, in questions like hairy black holes, isolated horizons
\cite{hairy} and entropy of black holes \cite{Myers}.

Besides, recently much attention has been deserved to nonlinear
electrodynamics in string theory, where solutions of Born-Infeld
equations represent states of D-branes \cite{Dbranas}.

In this paper we study the Einstein-Born-Infeld (EBI) solution that is
the nonlinear electromagnetic generalization of the R-N black hole. One
point of interest is the singularity that possesses this
Einstein-Born-Infeld solution.  Singularity that however, is not always
reached by photons or test particles, this depending on the balancing
between the three parameters mass, charge and the Born-Infeld parameter. 
Moreover, it turns out that the nonlinearity in the electromagnetic field
modifies the size of the horizon as well as the effective geometry seen
by the B-I photons.  Basic facts on B-I nonlinear electrodynamics are
sketched in Sec. 2. In Sec. 3 we present the EBI solution in appropriate
coordinates and study the relevant metric function. In Sec. 4 the
effective potentials are shown for test particles and photons. Some
conclusions are drafted in the last section.


\section{The Born-Infeld nonlinear electrodynamics}

For completeness we include in this section the basic facts on nonlinear
electrodynamics proposed by Born and Infeld \cite{Born} in the formalism
implemented by Pleba\~nski \cite{Pleban} for solutions of Petrov type D.
The Born-Infeld electrodynamics is derived from an action for the
gravitational electromagnetic field given by

\begin{equation}
S = \int{d^4x \sqrt{-g} \{ R (16\pi)^{-1}+L \} } ,
\end{equation} 

where $R$ denotes the scalar curvature, $g:= {\rm det} \vert g_{\mu \nu}
\vert$ and $L$, the electromagnetic part, is assumed to depend on the so
called structural function $K(P,\check Q)$ constructed from the
invariants of the electromagnetic field, $P $ and $\check Q$,

\begin{equation}
L=-{1 \over 2} P^{\mu \nu} F_{\mu \nu} +K(P,\check Q),
\end{equation}

where $F_{\mu \nu}$ corresponds to the intensity of electric field and
magnetic induction vectors ({\bf $E$} and {\bf $B$}) while the
antisymmetric tensor $P^{\mu \nu}$ corresponds to the electric induction
and the intensity of the magnetic field vectors ({\bf $D$} and {\bf
$H$}). $P$ and $\check Q$ are the two invariants of the tensor $P_{\mu
\nu}$: 

\begin{equation} P:={1 \over 4}P_{\mu \nu}P^{\mu \nu}, \check Q={1 \over
4}P_{\mu \nu}\check P^{\mu \nu}, 
\end{equation}

with $\check P_{\mu \nu} :=-{ \epsilon_{\mu \nu \rho \sigma} \over {2
\sqrt{-g}}}P^{\rho \sigma}$, $\epsilon_{\mu \nu \rho \sigma}$ is the
totaly antisymmetric Levi-Civita tensor.  The admissible structural
functions $K(P, \check Q)$ are constrained to fulfill the requirements of
the correspondence to the linear theory ($K=P+O(P^2, \check Q^2)$), the
parity conservation ($K(P, \check Q)=K(P,- \check Q)$), the positive
definitness of the energy density ($K_{,P} >0$) and the requirement of
the timelike nature of the energy flux vector ($P K_{,P}+ \check Q K_{,
\check Q}-K \ge 0$), where $K_{,P}:= \frac{\partial K }{\partial P}$ and
$K_{, \check Q}:= \frac{\partial K}{\partial \check Q}$.

Performing the variation of $S$ with respect to $g_{\mu \nu}$, $A_{\mu}$
and $P^{\mu \nu}$, the least action principle, $\delta S=0$, lead to the
dynamical equations, namely, Einstein equations, Maxwell-Faraday
equations and the material or constitutive equations (that relate
$F_{ab}$ with $P_{ab}$). 

Working in the null-tetrad formalism, the field structures to be studied
are given by the metric

\begin{equation} g=2e^1 \otimes e^2+2e^3 \otimes e^4, \quad e^1=\bar e^2
, \quad e^3=\bar e^3, \quad e^4=\bar e^4, \end{equation}

Accompanied by the two-form of the nonlinear electromagnetic field,
\begin{equation}
\omega={1 \over 2}(F_{ab}+\check P_{ab})e^a \wedge e^b,
\end{equation}
where $\check P_{ab} :=-{1 \over 2} \epsilon_{abcd}P^{cd}$ and
the tensor field $F_{ab}$ is determined through the material equations
\begin{equation}
F_{ab}=K_{,P}P_{ab}+K_{, \check Q} \check P_{ab}.
\label{mateq} 
\end{equation}

The closure condition of the two-form $\omega$ is equivalent to the
Maxwell and Faraday equations: 

\begin{equation}
d \omega =0 \rightarrow  \check F^{ab}{}_{;b}=0, \quad P^{ab}{}_{;b}=0.
\end{equation}

The system of equations for NLE is closed by the Einstein equations
\begin{equation}
R_{ab}-{1 \over 2} g_{ab} R=8 \pi E_{ab}, 
\end{equation}
where the energy-momentum tensor $E_{ab}$ is given by

\begin{equation} 4 \pi E_{ab}=K_{,P}(-P_{as}P_b{}^{s} + g_{ab}
P)+(PK_{,P}+ \check Q K_{,\check Q} -K)g_{ab}, \label{nletensor}
\end{equation}

the scalar curvature $R$ being
\begin{equation}
R=-8(PK_{,P}+ \check Q K_{,\check Q} -K).
\end{equation}

The Born-Infeld nonlinear electrodynamics is characterized by the
structural function

\begin{equation}
K=b^2(1-\sqrt{1-\frac{2P}{b^2}+ \frac{\check Q^2}{b^4}}),
\label{BIK}
\end{equation}

the parameter $b$ is the electric field at the origin ($r=0$). The
limiting transition $b \to \infty$ guarantees the correspondence to the
linear Maxwell theory with $K=P$. The structural function in (\ref{BIK})
is singled out among all possible structural functions by leading to a
single family of characteristic surfaces \cite{Boillat}.

For metrics of the Petrov type D one can always align the directions of
the real null vectors, $e^3, e^4$, along the Debever-Penrose vectors. 
Also the eigenvectors of $F_{ab}$ (and consequently the ones of $P_{ab}$) 
can be aligned in the directions of the Debever-Penrose vectors. These
alignments leave as nonvanishing components of $F_{ab}$ ($P_{ab}$),
$F_{12}$, $F_{34}$ ($P_{12}$, $P_{34}$).


\section{Born-Infeld generalization of
Reissner-Nor\-dstr\"om solution}

In this section some remarkable properties of the Born-Infeld (B-I)
generalization of the Reissner-Nordstr\"om (R-N) solution are given. 

This Einstein-Born-Infeld solution was presented in \cite{GSP}. However,
it was presented in terms of a not very friendly elliptic integral and in
coordinates from which is not clear how to obtain its linear
electromagnetic limit (R-N) neither its uncharged limit (Schwarzschild) 
(cf. Eqs. (4.11)-(4.16) in \cite{GSP}).  In the so-called canonical
coordinates ($t,r, \theta, \phi$) in which the R-N metric is usually given
\cite{Chandra} and writting the elliptic integral (which is of the first
kind) in terms of the Legendre's elliptic functions
$F(\beta,k):=\int^{\infty}_{\beta}{(1-k^2 \sin^2s)^ {-\frac{1}{2}}}ds$,
the solution is given by

\begin{eqnarray} ds^2&=& - \psi dt^2+\psi^{-1}dr^2+r^2(d \theta^2 +
\sin^2{\theta}d \phi^2), \\ \psi&=& 1-\frac{2M}{r} -\frac{\lambda
r^2}{3}+ \frac{2}{3}b^2 r^2 (1- \sqrt{1+ \frac{Q^2}{b^2r^4}})+
\nonumber\\ && \frac{2Q^2}{3r}\sqrt{\frac{b}{Q}} F(\arccos{\{
\frac{br^2/Q-1}{br^2/Q+1}} \},\frac{1}{\sqrt{2}}), 
\label{BImetrfunc}
\end{eqnarray}

\begin{equation}
F_{rt}= Q (r^4+ \frac{Q^2}{b^2})^{- \frac{1}{2}},
\label{FrtBI}
\end{equation} 

Where $G=c=1$, $M$ is the mass parameter, $Q$ is the electric charge
(both in lenght units) and $b$ is the Born-Infeld parameter which
corresponds to the magnitude of the electric field at $r=0$, given in
units of $[\rm{lenght}]^{-1}$. The solution includes the cosmological
constant $\lambda$, in such a manner that one of the limit cases is the
de Sitter solution (when $M=b=Q=0$). With the substitution $Q \to
\sqrt{Q^2+g^2}$, the solution includes the so-called magnetic charge $g$
(remind that the Born-Infeld theory has the freedom of the duality
rotations).  In this work we put $\lambda =0$ and $g=0$.

As we state before, in nonlinear electromagnetism, the role of the skew
symmetric field tensor $F_{\mu \nu}$ is now played by the tensor $P_{\mu
\nu}$; both are related through the material or constitutive equations
(\ref{mateq}). For the Born-Infeld R-N solution $P_{\mu \nu }$ is given by

\begin{equation}
P_{rt}= \frac{Q}{r^2} ,
\end{equation} 

The field $P_{rt}$ is singular at the origin while $F_{rt}$, Eq. (
\ref{FrtBI}), at the origin gives a finite value corresponding to the
magnitude of $b$. $F_{rt}$ corresponds to the gradient of a potential
$A_t$

\begin{equation}
A_t(r)= Q 
\int^{\infty}_{r}{(\frac{Q^2}{b^2} + s^4)^{- \frac{1}{2}} ds},
\end{equation}

The two nonvanishing components of the nonlinear electromagnetic
Born-Infeld field, Eq. (\ref{nletensor}), in this case are given by

\begin{eqnarray}
8 \pi E_{12}&=&2b^2[(1+\frac{Q^2}{b^2r^4})^{-\frac{1}{2}}-1], \nonumber\\
8 \pi E_{34}&=&2b^2[(1+\frac{Q^2}{b^2r^4})^{\frac{1}{2}}-1],
\label{bicomp} 
\end{eqnarray}

In the limit of large distances, $r \to \infty$, aymptotically it
corresponds to the R-N solution. Also when the B-I parameter goes to
infinity, $b \to \infty$, we recover the linear electromagnetic
(Einstein-Maxwell) R-N solution, with

\begin{eqnarray} ds^2&=& - \psi_{RN} dt^2+\psi_{RN}^{-1}dr^2+r^2(d
\theta^2 + \sin^2{\theta}d \phi^2), \\ \psi_{RN}&=& 1-\frac{2M}{r} +
\frac{Q^2}{r^2} \\ F_{rt}&=& \frac{Q}{r^2} , \end{eqnarray}

To obtain these limits from Eqs. (\ref{BImetrfunc})-(\ref{FrtBI}) we must
consider the integral expression for the Legendre elliptic function $F$

\begin{equation} F(\arccos{\{ \frac{br^2/Q-1}{br^2/Q+1}}
\},\frac{1}{\sqrt{2}})=2 \sqrt{\frac{Q}{b}}
\int^{\infty}_{r}{(\frac{Q^2}{b^2} + s^4)^{- \frac{1}{2}} ds}. 
\label{ellfunc} \end{equation}

In the limit $b \to \infty$ the fourth term of the metric function
$\psi$, Eq. (\ref{BImetrfunc}) is $- Q^2/3r^2$ while the last term
amounts to $4Q^2/3r^2$; the sum of both terms amounts to $Q^2/r^2$,
according with $\psi_{RN}$.  Clearly, in the limit of $Q=0$ the
Schwarzschild solution is obtained.  We observe that at $r=0$ the last
two terms in $\psi_{RN}$ diverge. On the other side, for $\psi$ we have
that Eq. (\ref{BImetrfunc}) can be written (with $\lambda=0$) as

\begin{eqnarray} \psi &= &1+ \frac{2}{3}b^2 r^2 (1- \sqrt{1+
\frac{Q^2}{b^2r^4}}) \nonumber\\ && + \frac{2}{r} \{-M+
\frac{Q^2}{3}\sqrt{\frac{b}{Q}} F(\arccos{\{ \frac{br^2/Q-1}{br^2/Q+1}}
\},\frac{1}{\sqrt{2}})\}, \end{eqnarray}

the second term when evaluated at $r=0$ is finite; the elliptic function
$F$ (cf. Eq. (\ref{ellfunc})) also is finite when evaluated at $r=0$.
Then, depending on the value of the term in curly brackets, we have three
possible cases of behavior of $\psi$ in the vicinity of $r=0$: 

\noindent{(i)} $M > \frac{Q^2}{3}\sqrt{\frac{b}{Q}}F(\arccos{\{
\frac{br^2/Q-1}{br^2/Q+1}} \},\frac{1}{\sqrt{2}}), \quad$ $\psi$ diverges
as $- \infty$ at $r \to 0$.

\noindent{(ii)} $M < \frac{Q^2}{3}\sqrt{\frac{b}{Q}}F(\arccos{\{
\frac{br^2/Q-1}{br^2/Q+1}} \},\frac{1}{\sqrt{2}}), \quad$ $\psi$ diverges
as $ + \infty$ at $r \to 0$.

\noindent{(iii)} $M = \frac{Q^2}{3}\sqrt{\frac{b}{Q}}F(\arccos{\{
\frac{br^2/Q-1}{br^2/Q+1}} \},\frac{1}{\sqrt{2}}), \quad$ $\psi$ does not
diverge at $r \to 0$. 

In spite that in case (iii) $\psi$ is regular in all the range, the
invariants diverge at $r=0$, that is the conclusion from the analysis of
the Weyl scalar $\Psi_2$.  Since the solution is of type D, the only
nonvanishing Weyl scalar is $\Psi_2$,

\begin{equation} \Psi_2= \frac{M}{r^3} - \frac{Q^2r^3}{6} \partial_r
(\partial_r (\frac{1}{r^2}{\int^{\infty}_{r}{(s^2+ \sqrt{s^4+
\frac{Q^2}{b^2}})^{-1}ds}})).  \label{WeylBI} \end{equation}

The invariants of this solution depend on $\Psi_2^2$, then at $r=0$ there
is a singularity at leats of the order $1/r^6$, comig from the first
(mass) term, similar to the case of Schwarzschild and
Reissner-Nordstr\"om.  Furthermore, the second term in Eq. (\ref
{WeylBI}), due to the electromagnetic field, also diverges at $r=0$. Then
in this spacetime there is solely one singularity at $r=0$.
  
The zeros of the metric function $\psi$ indicate the existence of
coordinate singularities, which can be eliminated by a change of
coordinates. The entire analytic extension may consist of several of such
patches, overlapping in finite regions. The procedure to follow to
determine these analytic continuations is given by Graves and Brill in
\cite{Graves} (see also \cite{Chandra} for analytical extension of R-N
metric). It consists in determining a simultaneous transformation of $r$
and $t$ to $[u(r,t), v(r ,t)]$ in terms of which the light cones are
lines with slopes $ \pm 1$, then in terms of these $(u,v)$ coordinates
the metric is regular. Such procedure can be used whenever the
singularity in the metric function $\psi$ is a zero of the first order.
That turns out to be the case for the metric function $\psi$ Eq. 
(\ref{BImetrfunc}), since when the lower limit in the integral expression
(\ref{ellfunc}) is fixed, then the Legendre elliptic function is a
constant.

The zeros of the metric function $\psi$ have to be localized numerically.
To this end we illustrate for several values of the parameters the
behavior of the metric function $\psi$.  In these plots $\psi$ is in
terms of the adimensional coordinate $u=r/M$, as

\begin{eqnarray} \psi &= &1-\frac{2}{u} + \frac{2}{3}(bM)^2 u^2 (1-
\sqrt{1+ \frac{\alpha^2}{(bM)^2u^4}})+ \nonumber\\ && \frac{2
\alpha^2}{3u}\sqrt{\frac{bM}{\alpha}} F(\arccos{\{ \frac{bMu^2/
\alpha-1}{bMu^2/ \alpha+1}} \},\frac{1}{\sqrt{2}}), \end{eqnarray}

\begin{figure}[htbp]
\centering \epsfig{file=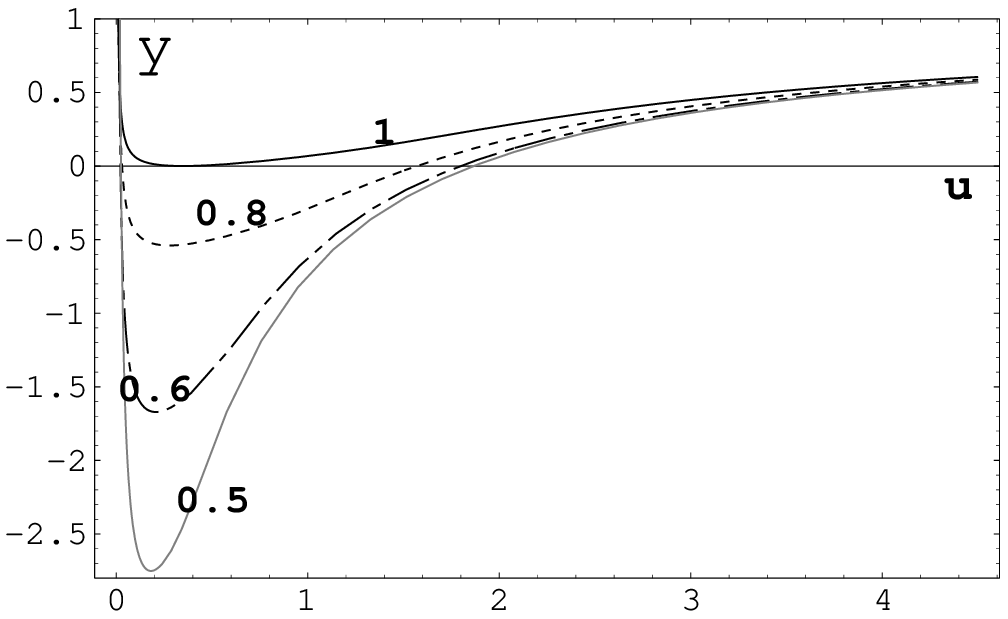, width=7.5cm}
\end{figure}
{\small \noindent{Fig. 1a.} It is shown the shape of the metric function
$\psi$
vs.  the coordinate $u=r/M$. The location of $r_h=u_hM$, with
$\psi(r_h)=0$, is the position of the horizon. The values of $b$ were
chosen such that at $u \to 0$, $\psi \to + \infty$. The value of
$\alpha=Q/M$ is printed on each curve, the corresponding $b$'s are:
$(\alpha,b)$, $(0.5,4.5/M)$, $(0.6, 2.5)$, $(0.8,1.03)$, $(1,0.5225)$.}


\bigskip

\begin{figure}[htbp]
\centering \epsfig{file=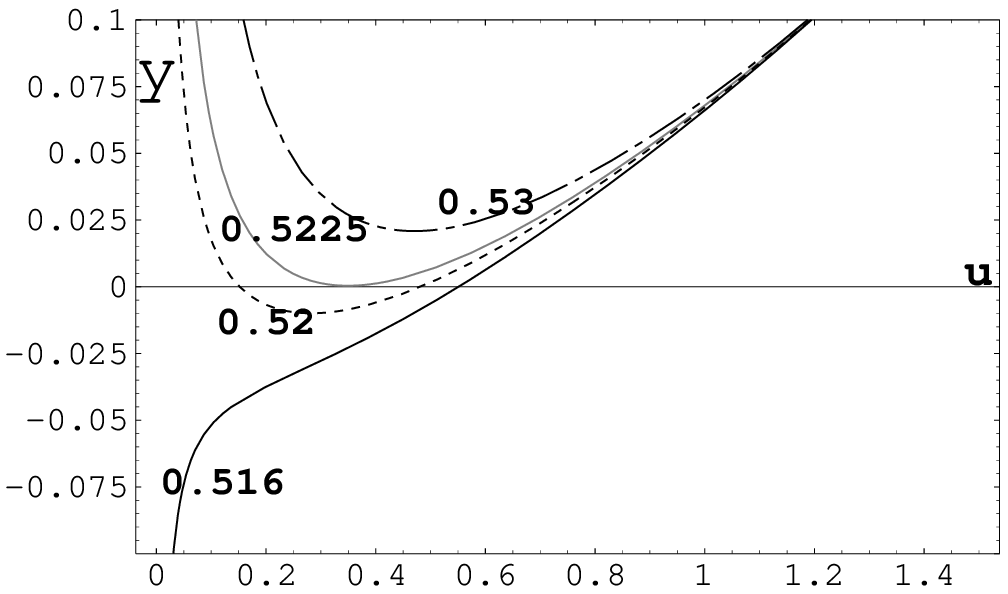, width=7.5cm}
\end{figure}
{\small \noindent{Fig. 1b.} It is shown the behavior of the metric
function
$\psi$ for the extreme case $(Q=M)$ with different values of $b$. For $b
> 0.5225/M$ the function $\psi \ge 0$ and the metric is regular
everywhere except at the origin.  A naked singularity is present. If $b
\le 0.516/M$, the Schwarzschild behavior dominates.}


\bigskip

In the plots we shall use, instead of $Q$, the constant $\alpha$,
$\alpha^2= \frac{Q^2}{M^2}$, this constant measures the ratio between the
charge and mass of the body that produces the field. $b$ and $M$ can vary
independently and $(bM)$ is adimensional. We identify the position of the
horizon as the value for which $g_{tt}$ ($= \psi$) is zero.  From the
graphics for $\psi$ (figs. 1a, 1b) we see that the zeros of this function
can be one, two or not occur at all. 
 
For $\alpha >1$ (hyperextreme case, $Q > M$) $\psi=0$ has no zeros, as in
the R-N case .  For small $\alpha$'s ($\alpha < 0.4$) the behavior
resembles the Schwarzschild one, independently of the value of $b$. In
this case the gravitational field due to the mass $M$ overwhelms the
(linear or nonlinear) electromagnetic field. 

For values of $0.5 < \alpha < 0.9$, the metric function resembles
Schwarzschild if $b < 0.7$ (weak electromagnetic field), see Fig. 1a. 
For these values, $0.5 < \alpha < 0.9$, the magnitude of $bM$ can be
adjusted in order that $\psi$ grews to $+ \inf ty$ near $r=0$. In these
cases $\psi$ has two zeros. In Fig. 1a it is also apparent that as
greater is $\alpha$, smaller is the value of $b$ that is sufficient to
defeat the gravitational attraction at $r=0$. The reason is that as
$\alpha$ grows, then the mass diminish in relation to the charge, and a
smaller nonlinear electromagnetic field (smaller $b$)  overwhelms the
gravitational one.

For $\alpha$, $\alpha > 0.5$ and $b >
4.5/M$, $\psi$ goes to infinity ($+ \infty$) at $r \to 0$, resembling a
soliton like behavior.  Furthermore, as greater is $b$, the point where
$\psi$ becomes zero, is nearer the origin, i.e. the size of the horizon
stretches (see fig. 1b).

Special attention deserves the case $\alpha=1$, that corresponds in the
black hole terminology to the extreme case ($Q=M$). In fig. 1b., for
$\alpha =1$, the function $\psi$ shows a very sensitive behavior with
respect to the value of $b$ in the neighborhood of $b=0.5/M$. In the
vicinity of this value three cases can occur: one horizon, two horizons
or not horizon at all.  When $\alpha=1$ and $b = 0.5225/M$ the metric
function $\psi$ possesses one horizon and goes to $+ \infty$ near $r=0$.
If $\alpha=1$ and $b \ge 0.53/M$ then $\psi$ has no zeros and presents a
naked singularity. For $\alpha=1, b < 0.516 M$, gravity dominates and the
behavior is Schwarzschild-like.

We pass now to describe some features of the trajectories of test
particles.


\section{Trajectories of test particles, photons and gravitons}
 
As in the Schwarzschild and R-N case, the timelike and null geodesics in
the equatorial plane ($\theta = \frac{\pi}{2}$) of the B-I spacetime
reduces to solving a problem of ordinary one-dimensional motion in an
effective potential.

\subsection{Charged and uncharged test particles}

Here we examine the law of motion of test particles of mass $\mu$ and
charge $\epsilon$. The geodesic equation is

\begin{equation} \frac{d^2 x^{\nu}}{d \tau^2}+\Gamma_{\alpha
\beta}{}^{\nu} \frac{dx^{\alpha}}{d \tau} \frac{dx^{\beta}}{d \tau}= -
\frac{\epsilon}{\mu} F_{\sigma}{}^{\nu} \frac{dx^{\sigma}}{d \tau},
\label{geodeq} \end{equation}

The test particle has two conserved quantities
at least, corresponding to the two Killing
vectors $\partial_{t}$ and $\partial_{\phi}$,
i.e. the energy of the particle and its
angular momentum, $\tilde l$. If the test
particle is charged, then the energy involves
an electromagnetic part that arises from the
right hand side of Eq. (\ref{geodeq}),
corresponding to the only nonvanishing
component of the electromagnetic field,
$F_{rt}$. The geodesic equation for the $t$
coordinate amounts to

\begin{equation}
\ddot{t}+\psi^{-1}(\psi_{,r})\dot{r}
\dot{t}=\frac{\epsilon}{\mu}\frac{F_{rt}}{\psi}
\dot{r}, \end{equation}

where dot indicates the derivative with
respect to an affine parameter.  This equation
can be integrated for $\dot{t}$, obtaining,

\begin{eqnarray} \dot{t} \psi &=& E +
\frac{\epsilon}{\mu}
\int^{\infty}_{r}{F_{rt}dr}, \nonumber\\
             &=& E + \frac{\epsilon Q}{\mu}
\int^{\infty}_{r}{(r'^4+\frac{Q^2}{b^2})^{-\frac{1}{2}}dr'}
\nonumber\\ &=& E + \frac{\epsilon
Q}{\mu}\sqrt{\frac{b}{Q}}\frac{1}{2}F(\arccos{\{
\frac{r^2-Q/b}{r^2+Q/b}}
\},\frac{1}{\sqrt{2}}).  \label{tdot}
\end{eqnarray}

where $E$ is the energy at infinity of the
uncharged test particle.  On the other hand,
from the line element for the timelike
geodesics we have,

\begin{equation} 1= \psi \dot{t}^2- \psi^{-1}
\dot{r}^2 -r^2 \dot{\phi}^2, \end{equation}

Substituting ${\tilde l}=g_{\phi
\phi}\dot{\phi}$, and $\dot{t}$ from Eq. 
(\ref{tdot}), we obtain

\begin{equation} \dot{r}^2 + \psi
(\frac{{\tilde l}^2}{r^2}+1) - [ E +
\frac{\epsilon Q}{2 \mu} \sqrt{\frac{b}{Q}}
F(\arccos{\{ \frac{r^2-Q/b}{r^2+Q/b}}
\},\frac{1}{\sqrt{2}})]^2 =0, \end{equation}

Comparing with $ \frac{1}{2} \dot{r}^2
+V_{eff}(E,{\tilde l},r)=0$, we get the
effective potential, which depends
nontrivially on $E$ as well as on ${\tilde
l}$, for a charged test particle.  The plots
of $V_{eff}$ are in terms of the adimensional
coordinate $u$, and $l= {\tilde l}/M$,
\begin{equation} V_{eff}(u)= \frac{\psi}{2} (\frac{{\it l}^2}{u^2}+1) - [
E + \frac{\epsilon \alpha}{2 \mu} \sqrt{\frac{bM}{\alpha}} F(\arccos{\{
\frac{u^2-\frac{\alpha}{bM}}{u^2+ \frac{\alpha}{bM}}}
\},\frac{1}{\sqrt{2}})]^2/2, \end{equation}

\begin{figure}[htbp]
\centering \epsfig{file=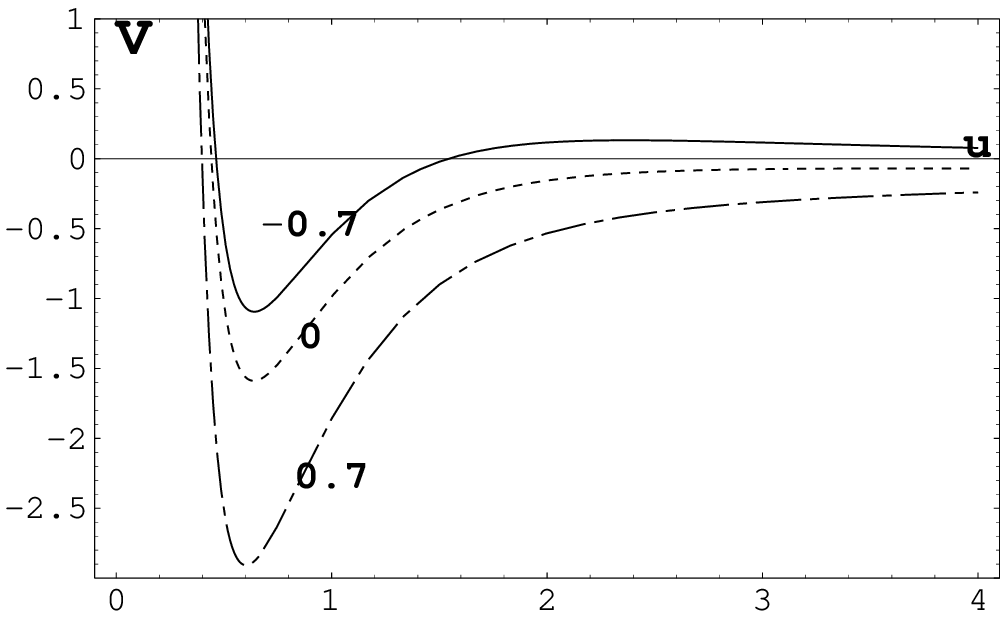, width=7.5cm}
\end{figure}
{\small 
\noindent{Fig. 2a.} The effective potential,
$V$, for test particles with different ratio
charge/mass, $\epsilon/ \mu$, as printed over
each plot, $\epsilon/ \mu= -0.7, 0, 0.7$. The
values of the constants are $E=1, l=3, \alpha=
0.9, b= 0.9/M$. The change in sign of the
charge results in a quantitative difference.
There are stable equilibrium positions that do
not depend much on the ratio $\epsilon/ \mu$. 
}


\begin{figure}[htbp]
\centering \epsfig{file=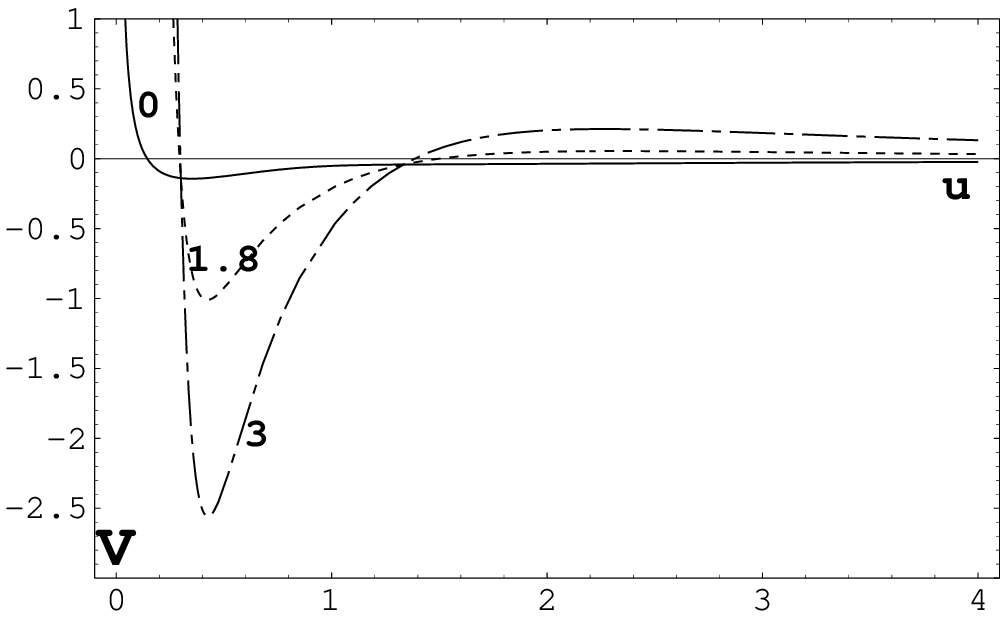, width=7.5cm}
\end{figure}
{\small 
\noindent{Fig. 2b.} The effective potential,
$V$, for test particles varying the angular
momentum, ($l= \tilde l /M$, where $\tilde l$
is the angular momentum of the test particle),
as printed, $l= 0, 1.8, 3$. The values of the
constants are $E=1, \epsilon / \mu = -1,
\alpha= 0.9, b=0.8/M$.  There are stable
equilibrium positions of lower energy for
greater $l$.
}


In Fig. 2a it is shown the effective potential
for different charge/mass,
$\frac{\epsilon}{\mu}$, values of the test
particle. The shape of the potential is the
same independently if the particle has
positive, negative or not charge at all. There
is an a ttractive region with stable
equilibrium positions. Fig. 2b corresponds to
the variation of the angular momentum of the
test particle. 

\begin{figure}[htbp]
\centering \epsfig{file=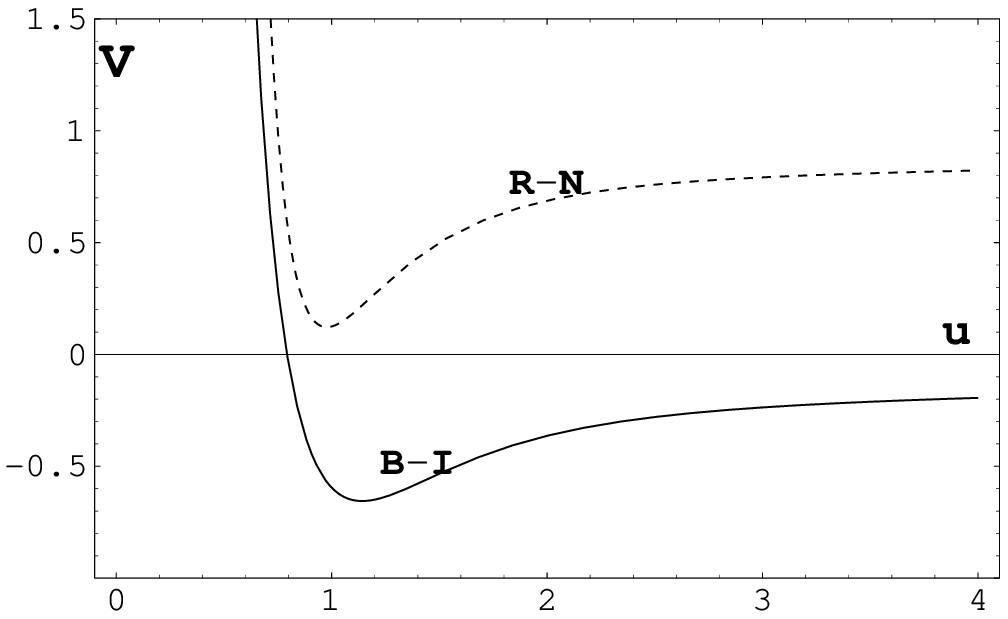, width=7.5cm}
\end{figure}
{\small 
\noindent{Fig. 3} In this figure the effective
potential as felt by a test particle in R-N
and in B-I is compared in the extreme case
($M=Q$). The constants are $\epsilon/ \mu=0.5,
E=1, l=3$ and for B-I, $b=0.8/M$. 
}


\bigskip
In Fig. 3 the comparison between the
Born-Infeld and Reissner-Nordstr\"om effective
potentials is shown in the extreme case
($Q=M$).  We remind that in R-N spacetime, test
particles do not reach the singularity at
$r=0$, while photons always do.  In B-I
spacetime, both behaviors are possible: to
reach or not the singularity, depending on the
value of the B-I parameter $b$. In particular,
test particles can not scape the singularity
if the metric function $\psi$ goes to $ -
\infty$ when $r \to 0$.

\subsection{Photons and Gravitons}

In nonlinear electromagnetism photons do not propagate along null
geodesics of the background geometry. Instead, photons propagate along
null geodesics of an effective geometry which depends on the nonlinear
theory (see \cite{Pleban}). Recently Novello {\ it et al} have presented
studies on these lines \cite{Novello} and also Gibbons has addressed this
topic in the context of string theory \cite{Gibbons2}. 

The discontinuities of the field propagate obeying the equation for the
characteristic surfaces, what in ordinary optics are the so called
``eikonal equations". For a curved spacetime the equation for the
characteristic surfaces is

\begin{equation} g^{\mu \nu}S_{,\mu}S_{,\nu}=0.  \label{null}
\end{equation}

The corresponding trajectories of the ``rays" are the null geodesics. 
When nonlinear electrodynamics is involved, the corresponding equation is
(cf. Eq. (10.106) in \cite{Pleban})

\begin{equation} (g^{\mu \nu}+ \frac{4 \pi}{b^2}E^{\mu
\nu})S_{,\mu}S_{,\nu}= \gamma^{\mu \nu}S_{,\mu}S_{,\nu}=0,
\label{nlechar} \end{equation}

Eq.(\ref{nlechar}) shows that it is the energy momentum density, $E_{\mu
\nu}$ (Eq. (\ref{nletensor})), of the nonlinear field which is the
essential cause why these surfaces are in general not null surfaces
obeying Eq. (\ref{null}). Then when one conside rs the effective
potential corresponding to photons, instead of using the elements of
$g_{\mu \nu}$, we must use the ones of the effective metric $\gamma_{\mu
\nu}$. Eq. (\ref{null}) then governs the propagation of the gravitational
discontinuities, while Eq.(\ref{nlechar}) governs the propagation of
nonlinear electromagnetic discontinuities. Those surfaces are locally
normal surfaces to the trajectories of gravitons and photons,
respectively. The linear (Maxwell) case is obtained when $b \to \infty$: 
the second term on the left hand side of Eq. (\ref{nlechar}) vanishes and
Eq. (\ref{null}) is recovered.

The trajectories corresponding to null rays are obtained from the line
element

\begin{equation}
0 = \psi \dot{t}^2- \psi^{-1} \dot{r}^2 -r^2 \dot{\phi}^2,
\end{equation}

Substituting $E= -g_{tt} \dot{t}= \psi \dot{t} $ and $\dot{\phi}={\tilde
l}/r^2$ and comparing with $\dot{r}+V_{\rm g}=0$ we have

\begin{equation}
V_{\rm g}=-\frac{g_{tt}l^2}{2u^2}=\frac{\psi l^2}{2u^2},
\label{gravtraj}
\end{equation}

while for the photons the corresponding effective potential must be
obtained, in analogous way, from the effective metric $\gamma_{\mu \nu}$,

\begin{equation}
V_{\rm ph}= \frac{\psi l^2}{2u^2} (1+\frac{\alpha^2}{(bM)^2u^4})^{-1},
\label{phpot}
\end{equation}

The effective potential for photons is the one for gravitons but
modulated by the factor
$(1+\frac{\alpha^2}{(bM)^2u^4})^{-1}=(1+\frac{Q^2}{b^2r^4})^{-1}$. This
factor goes to one in the linear case ($b \to \infty$) and we recover
Eq. (\ref{gravtraj}).

Eq. (\ref{phpot})can be written as
\begin{eqnarray} V_{\rm ph}(r)&=& \frac{{\tilde l}^2}{2r^2}
(1+\frac{Q^2}{b^2r^4})^{-1},\nonumber\\ &=&\frac{{\tilde l}^2}{2} \{
r^2-2Mr+\frac{2b^2}{3}[r^4 - \sqrt{r^8+\frac{Q^2r^4}{b^2}}] \nonumber\\
&&+\frac{2}{3}\sqrt{\frac{b}{Q}}Q^2 r F(\arccos{\{
\frac{br^2/Q-1}{br^2/Q+1}} \},\frac{1}{\sqrt{2}})\}, \label{phpot1}
\end{eqnarray}

from the last expression it is evident that at
$r=0$ the effective potential for the photon
is zero, $V_{\rm ph}(0)=0$. In contrast, when
approaching the origin, the graviton potential
$V_{\rm g}$ grows to plus infinity ($ +
\infty$), resembling a repulsive particle-like
potential (in analogy with R-N). However, for
the regions where the radial coordinate is $r
< r_{h}$, with $\psi(r_h)=0$, we must keep in
mind that $r$ is actually a timelike
coordinate, hence, it could be misleading to
think on region $r < r_{h}$ as the {\it
interior} of the black hole (excepting cases
of double root when $\alpha=1$).

\begin{figure}[htbp]
\centering \epsfig{file=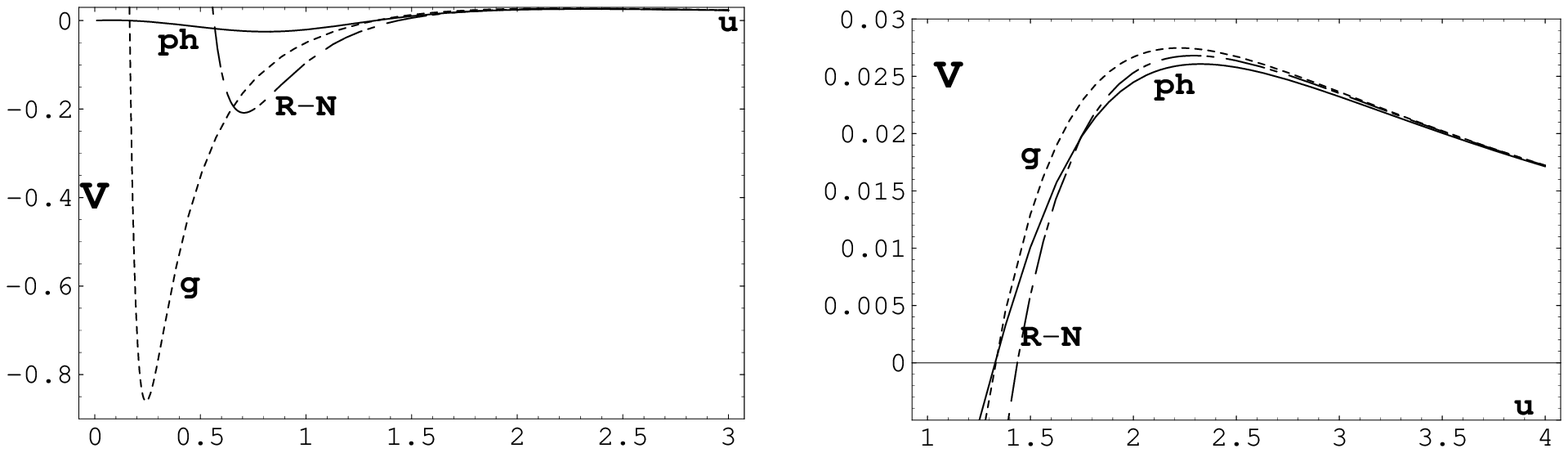, width=13cm}
\end{figure}
{\small 
\noindent{Figs. 4} It is shown the effective
potential corresponding to null geodesics in
R-N and B-I. In the latter case it is
different the effective potential felt by a
photon (ph) and the one felt by a graviton (g,
null geodesics). The points where $V(u)=0$
corresponds to the horizon and it is shared by
photon and graviton in the B-I case; it is
closer to the origin than in the R-N case (the
B-I field srinks the horizon). The values of
the constants and parameters are: $\alpha=0.9,
b= 0.75/M, l=1$. The plots correspond to
different ranges of $u$.
}


The different behavior is compared in Fig. 4
for both effective potentials and also
compared with the Reissner-Nordstr\"om null
trajectories for the same value of the
charge-mass relation, $Q/M=.9$, and the same
angular momentum ($l=1$). For this plot the
Born-Infeld parameter is $b=.75/M$.  Of course
gravitons interact with the B-I field but just
through the effect of B-I field on the
spacetime geometry.

\section{Conclusions}

We have studied the nonlinear electromagnetic generalization
(Born-Infeld) of the Reissner-Nordstr\"om metric. It describes a
nonsingular, asymptotically flat spacetime outside a regular event
horizon.  The solution is interesting because a straight comparison with
R-N can be established. The effect of increasing the Born-Infeld
parameter $b$ is to srink the size of the horizon.  Solutions without or
with one or two horizons are possible, depending on the values of the
parameters. Another feature of this solution is that, while for R-N all
photons reach the singularity at $r=0$, for the Born-Infeld
generalization, some of them skip the singularity at $r=0$.

Asymptotically this solution is a Reissner-Nordstr\"om one and then the
global charges defined at spatial infinity such as ADM mass and electric
charge are the global parameters that describe this solution. However the
mass and electric charge do not determine completely the solution. For
each solution characterized by the parameters $(M,Q)$, there exist an
infinite number of solutions with different Born-Infeld parameter $b$
with a different behavior near the horizon (see Fig. 1b, for instance). 
In this sense this is a colored black hole, with an Abelian (Born-Infeld) 
hair. Further investigation is needed to determine if the solution is
unstable.

In spite that the geometry for photons and gravitons is not the same,
both share the same horizon ( see Fig. 4, same $\psi(u)=0$). This feature
confirms for black holes, the conjecture claimed by Gibbons
\cite{Gibbons2} in the string context: {\it if the closed string metric
is static and the Born-Infeld field is pure electric or pure magnetic,
then the open string metric can not have a non-singular event horizon
distinct from the one given by the closed string metric}; it considering
that gravitons obey the closed string action while photons the open
string one, and that the solution studied here has a pure electric B-I
field. 

\section{Acknowledgments}

This work is partially supported by CONACyT (M\'exico), under project
32086-E

\end{document}